# Personalized Oncology: Feasibility of Evaluating Treatment Effects for Individual Patients

Lydia Jang,[1*] Stefan Konigorski[2,3*]

[1] Institute of Public Health, Charité – Universitätsmedizin Berlin
[2] Digital Health – Machine Learning Research Group, Hasso Plattner Institute for Digital Engineering, Potsdam, Germany
[3] Hasso Plattner Institute for Digital Health at Mount Sinai, Icahn School of Medicine at Mount Sinai, New York, NY, USA

*Corresponding authors: lydia.jang@charite.de, stefan.konigorski@hpi.de

## Abstract

The effectiveness of personalized oncology treatments ultimately depends on whether outcomes can be causally attributed to the treatment. Advances in precision oncology have improved molecular profiling of individuals, and tailored therapies have led to more effective treatments for select patient groups. However, treatment responses still vary among individuals. As cancer is a heterogeneous and dynamic disease with varying treatment outcomes across different molecular types and resistance mechanisms, it requires customized approaches to identify cause-and-effect relationships. N-of-1 trials, or single-subject clinical trials, are designed to evaluate individual treatment effects. Several works have described different causal frameworks to identify treatment effects in N-of-1 trials, yet whether these approaches can be extended to single-cancer patient settings remains unclear. To explore this possibility, a longitudinal dataset from a single metastatic cancer patient with adaptively chosen treatments was considered. The dataset consisted of a detailed treatment plan as well as biomarker and lesion measurements recorded over time. After data processing, a treatment period with sufficient data points to conduct causal inference was selected. Under this setting, a causal framework was applied to define an estimand, identify causal relationships and assumptions, and calculate an individual-specific treatment effect using a time-varying g-formula. Through this application, we illustrate explicitly when and how causal treatment effects can be estimated in single-patient oncology settings. Our findings not only demonstrate the feasibility of applying causal methods in a single-cancer patient setting but also offer a blueprint for using causal methods across a broader spectrum of cancer types in individualized settings.

# 1   Introduction

Personalized treatments in oncology offer the potential to deliver the most effective therapy for each patient, provided that treatment effects can be causally identified. Various ways have been used to address this need, with each contributing to the advancement of tailored therapies in oncology. Precision oncology can be described as approaches in cancer treatment that tailor therapies to the molecular characteristics of an individual patient's tumor rather than their morphology.[1] As one approach, treatments that target specific tumor characteristics have been developed, including a well-known example of HER-2 therapies like trastuzumab for breast cancer.[2] The combination of multi-omics, including genomics, transcriptomics, proteomics, epigenomics, metabolomics, and microbiomics, has offered an ability to provide a more thorough description of an individual's molecular profile.[3] Despite considerable developments to understand an individual's molecular dynamics and treatment response, cancer treatment remains a challenge. While molecular profiling methods have improved individual characterization, it does not always lead to consistent treatment outcomes. No suitable therapy may be available due to rare sequencing results, and even when treatment is administered, tumors can develop resistance mechanisms driven by factors not yet fully understood.[4]

These limitations in cancer treatment underline the necessity for methods other than molecular profiling to determine if a treatment is causally effective for each patient. N-of-1 trials, or single-subject clinical trials, have risen as a method for assessing individual treatment effects. Unlike population-level studies, which estimate average treatment effects across groups of patients, N-of-1 trials allow within-patient comparisons, giving a more personalized framework for evaluating outcomes.[5,6] Classical N-of-1 trials include a randomized, pre-specified framework and numerous treatment periods to allow adequate time for each treatment block.[7] These designs have typically been recommended for chronic conditions, which allow for stable follow-up and feasible washout periods, making repeated treatment comparisons possible.[5] In oncology, however, such designs are difficult to implement because of rapid disease progression and the challenge of establishing adequate washout periods.[8,9] Current N-of-1 studies in cancer have largely focused a priori on profiling the individual's characteristics at the start of the study and tailoring a treatment predicted to be most effective for the patient.[10,11] While distinct from traditional N-of-1 trials, these studies nevertheless highlight the importance of addressing the individual as the primary unit of observation, as opposed to population-level studies.

For instance, the I-PREDICT trial improved clinical outcomes by using tumor DNA sequencing to provide metastatic cancer patients with personalized therapies.[12] The WINTHER trial was the first multinational trial conducted across North America, Europe, and the Middle East to combine both genomic and transcriptomic data to guide individualized treatment for cancer patients.[13] A higher matching score, which calculated the degree of matching between the individual's molecular information and recommended treatment, was correlated with more optimal patient outcomes.[13] These patient-centric models, which are standardized to the individual and depart from the classic drug-centric approaches,[8] highlight the importance of tailoring treatment recommendations to improve patient outcomes in cancer care. Whether individualized oncology studies can incorporate elements of N-of-1 trials to enable causal estimation of treatment effects remains an open question.

To obtain valid causal estimates, the specialized designs of N-of-1 trials require careful methodological considerations.[14,15] The simplest statistical method to test for an effect in an N-of-1 trial is a t-test, which assumes that outcomes are independent across time and that no carryover or time effects are present.[16] Various works have developed statistical frameworks to define estimands under specific assumptions to provide a basis for drawing causal effects in N-of-1 trials.[17-19] Daza introduced a counterfactual framework for causal analysis in N-of-1 trials by defining an estimand that accounts for autocorrelation, carryover, and time trends. Malenica et al. developed several causal estimands to allow valid inferences at any time throughout an N-of-1 trial.[18] Piccininni et al. proposed a framework for applying causal inference in N-of-1 trials using the U-CATE estimand, which represents the average treatment effect conditional on the individual's baseline characteristics.[19] The paper further outlined the assumptions required for using a simple mean difference as an estimator and described situations, such as time trends, carryover effects, or outcome–outcome effects, where more advanced methods like the time-varying g-formula are needed.[19] Together, these papers emphasize the need to clearly define estimands and explicitly state the assumptions underlying their use in order to support valid causal conclusions in N-of-1 trials.

Given the various methodologies proposed for applying causal methods to N-of-1 trials, this study is motivated by the question of whether causal inference methods may be used to analyze treatment effects in a single-patient cancer dataset. This study specifically addresses the feasibility of applying causal inference to estimate treatment effects in one metastatic breast cancer patient, where adaptive treatment decisions were made by biomarker levels, imaging results, and clinical assessments. Given the challenges of assessing individualized treatment responses in oncology, this study aims to determine whether causal relationships between treatments, biomarkers, and lesion size can be systematically modeled. Although this study focuses on a specific case, its approach is applicable to other observational settings with adaptive and dynamic treatment decisions. The insights developed here could also be useful in other settings where treatment decisions are personalized and time-varying, extending beyond oncology.

# 2  Case Study

## 2.1 Overview

The patient data analyzed in this study were obtained from the supplemental materials of Johnson et al.[20] The female patient was diagnosed with hormone receptor-positive, HER-2 normal breast cancer at the age of 64. Computed tomography (CT) and fluorodeoxyglucose-positron emission tomography (FDG-PET) imaging results confirmed widespread disease after a first round of treatment. The patient was enrolled in a longitudinal monitoring program with clinical data collected over three and a half years. Throughout this period, the patient underwent treatment phases in response to disease progression.

The present study focuses on three primary data sources from the original dataset, which includes a detailed treatment regimen dataset, a serum tumor protein biomarker dataset, and a radiologic imaging dataset. The treatment regimen dataset includes a total of eleven different drugs administered to the patient during the entire period of the study. Out of the eleven administered drugs, three were given to moderate aspects of therapy-induced toxicities[20] and were not considered disease-directed therapies in the present study. The patient also received a series of radiotherapy sessions. A detailed treatment record was documented throughout the entire study period, which included the administration date and dosage of each drug. To note, hydroxychloroquine did not have this information included in the record and was excluded from analysis due to this reason.

The primary study utilized three different types of serum tumor protein biomarkers, cancer antigen 15-3 (CA 15-3), cancer antigen 27-29 (CA 27-29), and carcinoembryonic antigen (CEA) levels, to monitor treatment response over time. The current study used this biomarker dataset as one source of outcome data, with measurements taken approximately every three weeks across 59 distinct time points. Of these, two time points included only CA 15-3 and CA 27-29 measurements. Radiologic lesion size assessments, based on CT and FDG-PET scans, provided a second source of outcome data and were performed at 18 time points, roughly every three months. At each time point for the lesion size assessments, 16 different types of lesion locations, ranging from lymph nodes, liver, and spleen, were identified and measured.

## 2.2 Exposure and Outcome Variables

In this study, the exposure variable was defined as any of the eight drugs listed in the treatment dataset, excluding those not considered disease-directed therapies. The outcome variables were considered as biomarker levels and lesion size. According to the original publication, treatment decisions were guided through the utilization of the tumor biomarkers, supplemented by lesion size images defined by the Response Evaluation Criteria in Solid Tumors (RECIST) criteria.[20]

The three types of blood biomarkers used in the original publication have been used extensively in clinical settings, particularly for breast cancer patients. CA 15-3 is a serum marker that detects the shed or soluble forms of the transmembrane glycoprotein Mucin-1 (MUC-1), which is often overexpressed in breast cancer, particularly during progression and metastasis.[21-23] CA 27-29 similarly targets a circulating epitope of MUC-1[23,24] and demonstrates comparable characteristics to CA 15-3 in patients with breast cancer.[25] CEA is a glycoprotein involved in cell adhesion.[26] While CEA expression is typically seen throughout fetal development, particularly in the gastrointestinal tract, high levels are regularly found in a variety of malignancies, most notably adenocarcinomas.[27] While these markers are widely used in clinical practice to monitor metastatic breast cancer,[28,29] they are not considered definitive indicators of treatment efficacy.[30] Nonetheless, given their consistency across the patient's treatment course and their temporal alignment with treatment phases, they serve as reasonable proxies to evaluate treatment response in this single-patient setting.

In clinical practice, treatment decisions are generally not made based only on biomarker levels and is supplemented with lesion data,[28] and this reflects the process that the clinicians followed when changing the course of treatments. This study also applies a similar rationale when looking at treatment effects and considers both sources of outcome data to analyze treatment effects. The serum biomarker levels are used as the primary outcome in this study, as there are more data points to consider the temporal relationships of the treatment effects. Lesion size is considered as a secondary outcome when evaluating treatment effects. In the context of this study, both biomarker levels and lesion data are considered to be appropriate proxies to observe treatment effects.

## 2.3 Treatment Definitions

The treatment phases described in the original publication would not be sufficient for causal analysis, as they involved adaptive treatment decisions and multiple drugs that were administered at the same time. For causal inference in longitudinal datasets, it is necessary to have data where the exposure is consistent over time in order to observe outcome differences under different treatment conditions (e.g., Treatment A vs. Treatment B).[31] Although the dataset provided a rare and rich longitudinal setting for a single patient to potentially observe treatment effects under different treatment conditions, further data exploration was needed to determine its suitability for analysis.

As the treatment regimen dataset included exact administration dates, it was possible to link which drug was given to each biomarker time point. For each biomarker time point, the drug variables were assigned a value of 1 if the drug was administered at least once between the previous measurement day and the current measurement day. If not, a value of 0 was assigned. This process was applied across all time points for all drugs. The resulting biomarker dataset included additional columns for each drug with a binary indicator, as well as a *treatment combination* column that summarized all drugs given at each time point.

Additional steps were taken to determine if the *treatment combination* column contained the same type of combination for at least three consecutive time points in order to assess possible trends in biomarker levels. Combinations appearing in fewer than three consecutive time points were excluded, and this process resulted in four distinct treatment combinations, shown in Figure 1.

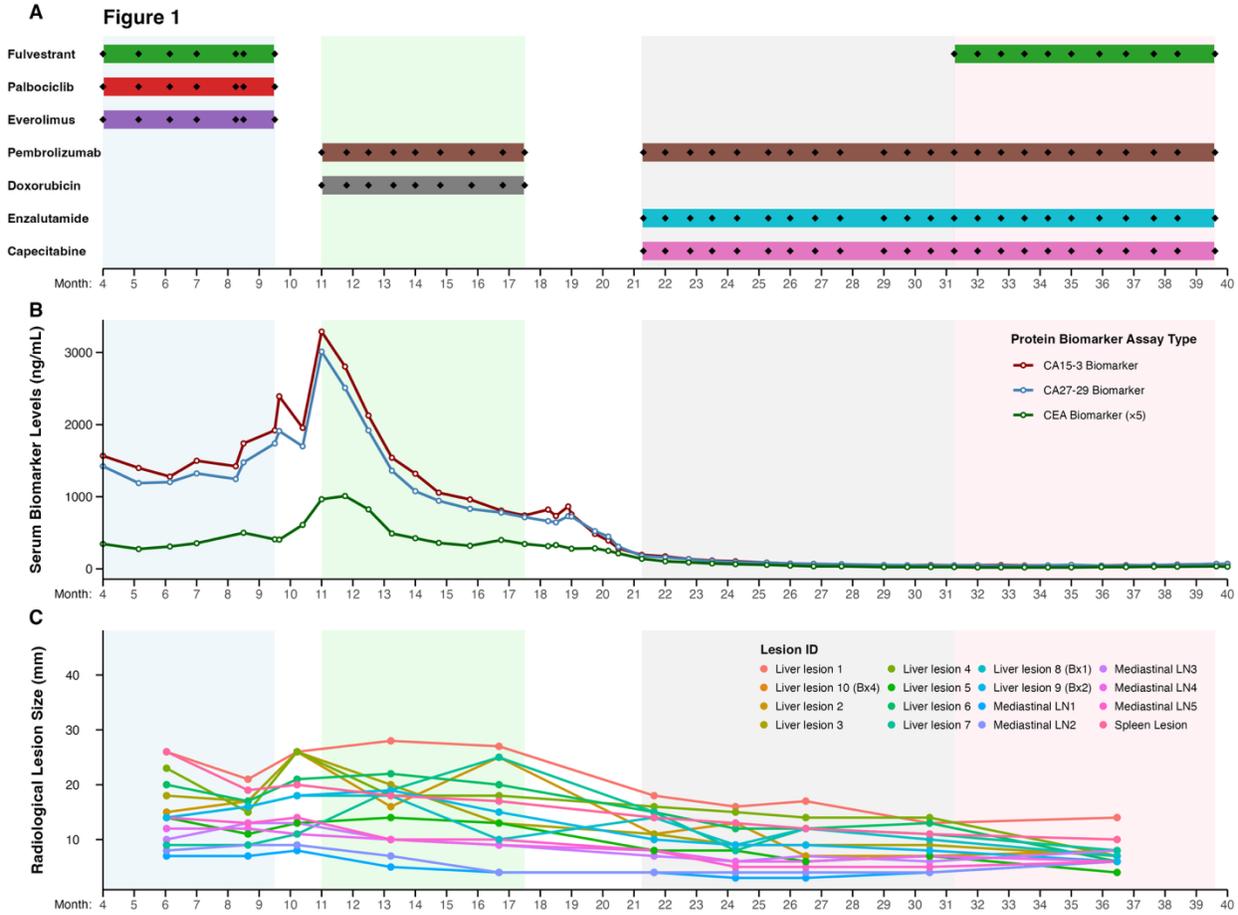

***Figure 1:*** *Trajectories of treatments, biomarkers, and lesion size after data preprocessing.*
*The layout of this figure is informed by Figure 2 in Johnson et al.,[20] but was redrawn following data preprocessing. The top panel, Figure 1A, shows the four treatment combinations that were identified after data exploration. Each row indicates a separate treatment, and the bullet points show when biomarker readings were obtained. The first combination consists of fulvestrant, palbociclib, and everolimus, which are marked from Month 4 to Month 9.5 with seven consecutive measures. The second combination consists of pembrolizumab and doxorubicin, administered from Month 11 to Month 17.5, with nine biomarker time points. The third combination, with pembrolizumab, enzalutamide, and capecitabine, included 12 consecutive time points. Finally, fulvestrant was added to this combination, resulting in a fourth combination, with 11 additional measurements. Figure 1B shows biomarker measurement trends across the shown study period, and Figure 1C shows lesion size trends.*

The dataset's suitability for causal analysis was initially uncertain because of the patient's adaptive treatment schedule. After data processing, the treatment periods with enough data points confirmed their potential for further analysis. To determine which treatment combinations could be compared to make meaningful comparisons, a potential option was to compare the first and second treatment combinations, contrasting the effects of a combination of three drugs (fulvestrant, palbociclib, everolimus) with a combination of two drugs (pembrolizumab, doxorubicin). According to the publication, Month 10 was marked as progressive disease under the RECIST criteria, at which point pembrolizumab and doxorubicin were introduced due to the worsening disease. The rising biomarker levels from Figure 1B around this time further confirmed worsening disease. This raised a question whether stationarity, an assumption in N-of-1 trials where the underlying outcome remains stable over time,[19] was satisfied in this setting.

In contrast to the first and second periods, the third and fourth periods provided a more stable basis for analysis. In Figure 1, between Day 595 (Month 21.3) and Day 1110 (Month 39.6), across 23 time points, the same three treatments–pembrolizumab shown in brown, enzalutamide shown in blue, and capecitabine shown in pink–were given continuously, with the only change being fulvestrant shown in green, added on Day 875, which coincides with around Month 31. During this period, biomarker levels remained relatively constant, and no lesion outcomes were flagged according to the RECIST criteria. Moreover, the original publication stated that the addition of fulvestrant was not based on a sudden rise in biomarkers or worsening clinical symptoms, but rather on consistent activity of the estrogen receptor signaling pathway, which remained consistently high throughout the whole longitudinal program of the patient.[20] The addition of fulvestrant, according to the publication's rationale, supports the assumption that the addition of fulvestrant may be less confounded by biomarker or lesion size changes compared to earlier treatment changes. For these reasons, we focused our analysis on the third and fourth blocks.

## 2.4 Fulvestrant Addition Scenario

We reference the third and fourth treatment periods as the Fulvestrant Addition Scenario (FAS). The pre-fulvestrant period is approximately nine months (12 outcome measurement time points) to observe baseline treatments: pembrolizumab, enzalutamide, and capecitabine. Fulvestrant was introduced into the existing treatment mix at the 13th outcome measurement time point and continued for an additional 11 time points, corresponding to approximately eight months. During the post-fulvestrant period, fulvestrant was administered seven times at intervals of roughly 4–5 weeks. In summary, the pre-fulvestrant trajectory includes around nine months, and the post-fulvestrant trajectory is around eight months.

# 3 Estimands and Assumptions for Causal Inference in Personalized Oncology

## 3.1 Causal Estimand

Under the FAS, the target estimand is defined as the difference between the pre- and post-fulvestrant addition periods, which compares average biomarker levels when fulvestrant was administered versus when it was not. The U-CATE estimand, formalized by Piccininni et al,[19] represents the average treatment effect conditional on the individual-specific variable $U$. Baseline individual-level features that are not captured by measured covariates, such as exposure or outcome factors, are represented by the unmeasured variable $U$. In the current dataset, examples of $U$ may include prior treatments, comorbidities, or other baseline characteristics of the patient that affect all outcomes. By restricting the inference to a single patient's characteristics, conditioning on $U$ allows the estimand to be defined as an individual-specific causal effect.[19]

We formally depict the estimand as:

$$U\text{-}CATE_k(u) = \mathbb{E}(Y_k^{\bar{a}_k=\bar{1}_k} | U = u) - \mathbb{E}(Y_k^{\bar{a}_k=\bar{0}_k} | U = u) \qquad (1)$$

where $Y_k$ denotes the outcome at time $k$, with data collected at $k = 1, \ldots, t$ time points. The variable $\bar{a}_k$ is the exposure history up to time $k$, and $U$ as the individual's baseline characteristics. $\bar{1}_k$ is a vector of size $k$ with all elements equal to 1, whereas $\bar{0}_k$ is a vector of size $k$ with elements of 0.

Given the 23 observed time points during the FAS, the estimand $U\text{-}CATE_k(u)$ is defined as the difference between the expected counterfactual outcome $\mathbb{E}(Y_k^{\bar{a}_k=\bar{1}_k})$ if the individual with $U = u$ had always received fulvestrant and the expected counterfactual outcome $\mathbb{E}(Y_k^{\bar{a}_k=\bar{0}_k})$ if the same individual had never received fulvestrant.

The application of this estimand to our dataset relies on the standard causal inference assumptions of exchangeability, consistency, and positivity,[32] which are necessary to define counterfactual outcomes from observed data. In this setting, it is important to consider that an underlying treatment combination of pembrolizumab, enzalutamide, and capecitabine was consistently given across all 23 time points. We define $\bar{a}_k = \bar{1}_k$, which indicates that fulvestrant (500 mg) was administered at least once between the previous and subsequent biomarker measurement time points for all $k$, in addition to the underlying combination therapy. $\bar{a}_k = \bar{0}_k$ indicates a period in which fulvestrant was not given, which we consider the control period. Under the FAS, time points 1-12 correspond to $\bar{0}_k$ and time points 13-23 correspond to $\bar{1}_k$.

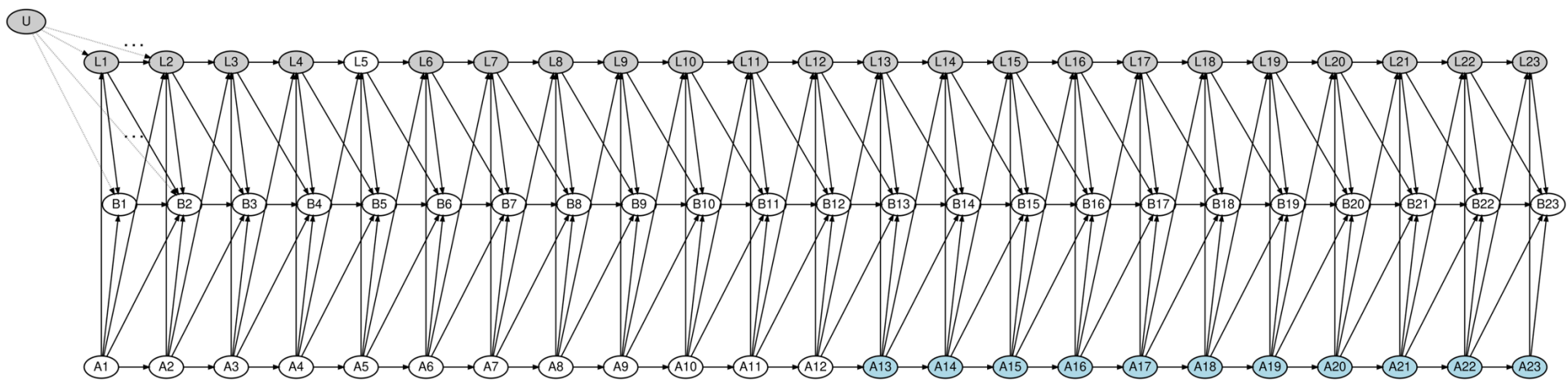

***Figure 2:*** *Directed Acyclic Graph for the fulvestrant addition scenario.*

## 3.2 Descriptions of the Directed Acyclic Graph

The relationships between the exposure and outcome variables are depicted in the Directed Acyclic Graph (DAG) shown in Figure 2. This DAG represents causal relationships between the exposures shown in A1-A12 nodes, the outcomes (biomarker levels as B1–B23 and lesion sizes as L1–L23), and the individual-specific variable $U$. Nodes represent variables, and directed edges represent causal relationships. Dashed arrows depict unobserved variables. To avoid clutter, we do not depict all unobserved arrows from $U$ to the outcome variables.

Throughout the treatment period shown in the DAG, a baseline regimen of three consistent therapies, pembrolizumab, enzalutamide, and capecitabine, was administered. The exposure nodes A1-A12 represent the pre-fulvestrant period, while the blue A13-A23 nodes represent the post-fulvestrant period, beginning on A13, the day fulvestrant was introduced.

As lesion size data were measured less frequently than biomarker levels, a last-observation-carried-forward (LOCF) approach was applied to align lesion size measurements with biomarker time points. In the DAG, lesion size nodes after LOCF are shown as white (observed), whereas grayed-out nodes denote latent (unmeasured) values. The L nodes directly represent lesion size, which also serves as a proxy for underlying disease progression.

The exposure node at every time point is assumed to affect both current outcomes (biomarker and lesion size), and its residual effects are assumed to also affect outcomes at the next time point. For both outcome variables, we assume that the current outcome is affected by the previous outcome. For instance, lesion size is not expected to reset at every time point, as the prior state is expected to affect the next measurement. Lesion size is assumed to influence the biomarker levels that follow, as larger tumors may release more circulating markers. This directionality does not apply to biomarkers, as biomarkers are outcomes of disease rather than drivers of progression. The individual-specific variable $U$, which represents unmeasured baseline characteristics or conditions unique to the individual that affect disease trajectory independent of treatment, is assumed to affect all outcomes.

## 3.3 Assumptions

Having defined the causal relationships with the DAG, we next address the key assumptions most relevant to this N-of-1 dataset. These assumptions represent fundamental challenges in N-of-1 designs, and we make them explicit to clarify under what conditions a causal effect of a treatment can be identified from the data. Specifically, we focus on the four assumptions that Piccininni et al[19] outlined when applying the U-CATE estimand in N-of-1 designs: carryover effects, outcome–outcome effects, time trends, and time-varying common causes, which, if present, represent violations of stationarity.

Carryover occurs when the effect of a treatment carries over and influences outcomes at subsequent time points. We look specifically into the period where fulvestrant is introduced in Figure 2, where exposure node A12 of the pre-fulvestrant period affects outcome node B13 of the post-fulvestrant period, indicating carryover to be present between the two periods. Outcome–outcome effects occur when a treatment changes an outcome at one time point, and that outcome then goes on to influence outcomes at subsequent time points. In our DAG, this is represented by arrows from outcome nodes to subsequent outcome nodes. Due to this, outcome–outcome effects are expected to be present during the FAS. A time trend exists when outcomes change over time, independent of treatment. As discussed in Section 3.2, lesion size serves as a proxy for disease progression. Lesion size may grow due to the natural progression of disease, apart from treatment effects, as tumors can develop at their own rate regardless of the treatment. In this dataset, using the LOCF approach to impute missing values for lesion size assumes that lesion size stays relatively stable between observed measurements, except at the points where a new measurement was recorded. We may expect a time trend whenever lesion size changes, as such changes are likely to cause biomarker levels to rise or fall accordingly. A similar consideration arises when considering the assumption for a time-varying common cause. Although we do consider lesion size to be a time-varying variable, it is treated as time-invariant at most time points, except at the points where a new observation was recorded. In other words, time variation exists but is not fully reflected because of the limited measurements and the imputation approach.

Based on these reasons, we assume carryover, outcome-outcome effects, time trends, and time-varying common causes of specific lesion size changes are present in this dataset. These violations imply that a t-test comparing pre- and post-fulvestrant addition is insufficient to state as the causal effect, and that methods like the g-formula are required to appropriately handle such violations.[19] As such, a time-varying g-formula was identified as an appropriate method to calculate the treatment effect under the FAS.

# 4 Causal Inference in Personalized Oncology

## 4.1 Data Generating Mechanism

We turn to a simpler representation of the DAG to formalize the structural equations for analysis. Figure 3 shows two time points from the DAG, the current and the one before. This choice allows the maximum number of causal relationships to be depicted, as the first time point alone would not capture lagged relationships. To understand the relationships influencing lesion size and biomarker levels given the exposure status, the red arrows highlight pathways affecting lesion size, while the blue arrows indicate pathways affecting biomarker levels.

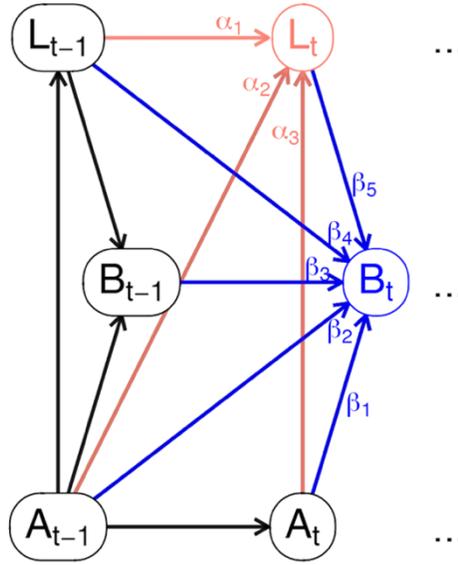

*Figure 3: Structural equation model with two selected time points for visualization. The model is representative of any t.*

Given the two outcomes of interest, we formalize the structural equations as follows,

$$B_t = \beta_0 + \beta_1 A_t + \beta_2 A_{t-1} + \beta_3 B_{t-1} + \beta_4 L_{t-1} + \beta_5 L_t + \varepsilon_B \tag{2}$$

$$L_t = \alpha_0 + \alpha_1 L_{t-1} + \alpha_2 A_{t-1} + \alpha_3 A_t + \varepsilon_L \tag{3}$$

where the biomarker level at time $t$, $B_t$, is determined by the current exposure $A_t$, the lagged exposure $A_{t-1}$, the previous biomarker level $B_{t-1}$, the lagged effects of the previous lesion size $L_{t-1}$, and the current lesion size $L_t$. Lesion size at time $t$, $L_t$, is determined by the previous lesion size $L_{t-1}$, the lagged exposure $A_{t-1}$, and the current exposure $A_t$. Error terms $\varepsilon_B$ and $\varepsilon_L$ are also indicated for both biomarkers and lesion size, representing the unexplained variation that is not accounted by the variables included in the structural equation. Not depicted in Figure 3, but we additionally consider the individual-specific variable $U$ to affect the outcomes.

We assume that the effects represented by the $\alpha$ and $\beta$ coefficients to be time-invariant, meaning they remain constant across all time points. For example, the coefficient $\beta_1$ on the path from $A_t$ to $B_t$ is the same effect as from $A_{t-1}$ to $B_{t-1}$, as well as for all other time points not explicitly depicted in Figure 3. Under this assumption, when fitting a linear regression to the dataset, the estimated $\beta_1$ reflects a single underlying effect that is averaged across repeated measures over time. This assumption enables pooling data from both pre- and post-fulvestrant periods to estimate a single effect. Without this assumption, each time point would need a separate coefficient, which is not feasible with limited data. Additionally, it is important to note that Figure 3 explains the data-generating mechanism of the FAS. The equations from (2) and (3) will be used to obtain outcomes under the different treatment assignments, "always treated" versus "never treated," to calculate the treatment effect.

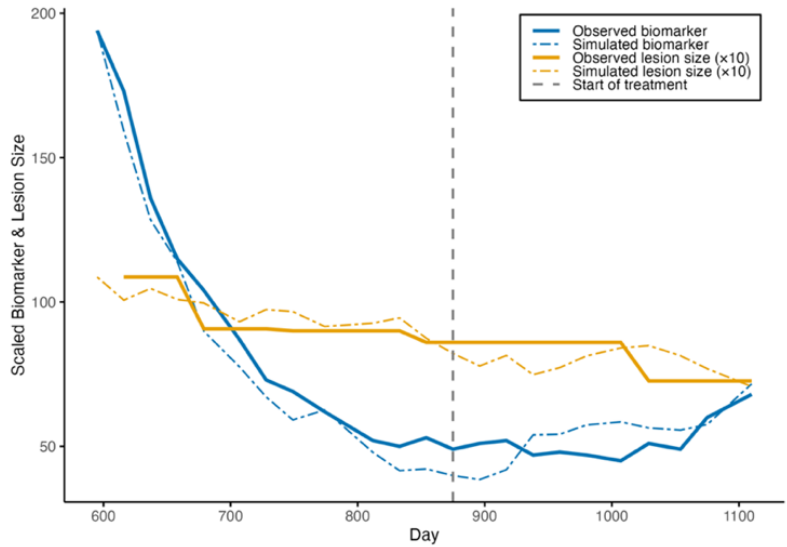

*Figure 4:* *Observed and simulated outcomes on biomarker levels and lesion size. Lesion size values were scaled by 10 for illustrative purposes.*

### 4.2 Model Validation

In this section, we use the equations in (2) and (3) for model validation before using them to estimate treatment effects. By comparing observed outcomes and simulated outcomes from the model, the validation determines if the model accurately reproduces biomarker and lesion size outcomes, and further supports its use for treatment effect estimation.

Figure 4 presents the observed and simulated outcomes obtained using structural equations defined in (2) and (3). For consistency, model validation and treatment effect estimation used CA 15-3 biomarker measurements among the three types of biomarkers. We assumed that the error terms from (2) and (3) were normally distributed. Using the `lm()` function in R, a linear regression model was first fit with biomarker level as the outcome and previous biomarker level, lesion size, previous lesion size, current exposure status, and previous exposure status as covariates. A second regression model was fit with lesion size as the outcome, using previous lesion size, current exposure status, and previous exposure status as covariates. The regression coefficients were used directly in the simulation, with initial biomarker and lesion size taken from the observed data, and subsequent values generated per the structural equations.

The model shows the general decline in biomarker levels and a stable trend in lesion size, with a mean squared error of 8.55 between simulated and observed outcomes. Although the model does not capture short-term intervals precisely, as seen in the slight underestimation and overestimation from Days 800-900 and 920-1000, the overall alignment in long-term trends indicates that the structural equations are sufficient for estimating causal effects.

### 4.3 Treatment Effect Estimation

To estimate the causal effect, outcomes were forward simulated using two regression models based on structural equations (2) and (3), with initial biomarker and lesion size measurements fixed from the observed dataset. Under the treatment scenario "always treated," the exposure status was set to 1 starting on Day 875, or the 13th time point, which was the day fulvestrant was actually given. Exposure status was set to 0 for all time points under the "never treated" treatment scenario. These two treatment scenarios were simulated across 500 repetitions using a g-computation Monte Carlo algorithm. The individual causal effect was calculated as the average mean differences for the "always treated" period and the "never treated" period, with parametric bootstrapping used to obtain the 95% confidence intervals. The point estimate was -38.1 (95% CI: -57.9, -18.4), indicating that biomarker levels were about 38.1 units lower in the "always treated" fulvestrant scenario compared to the "never treated" scenario. All analyses were conducted using RStudio v2025.05.1+513.

# 5 Discussion

This study explored the feasibility of using causal inference methods to estimate treatment effects in a dataset of a single metastatic breast cancer patient who received several treatments over time. The availability of a longitudinal dataset consisting of 59 biomarker measurements taken every 3-4 weeks, supplemented by radiological imaging results every three months, provided a unique opportunity to look into treatment effects in a single cancer patient over time. Such detailed data collection is rarely documented in oncology, where treatment monitoring is typically limited to imaging every few months.[33] The exact administration dates of each drug during the course of the study allowed the exposure and outcome to be temporally aligned. The four distinct treatment periods with sufficient time points provided a basis for meaningful comparisons and an appropriate data structure for conducting causal inference. Our main analysis, which focused on the FAS, allowed for a direct comparison of treatment periods where no gaps in treatment between the contrasts were present. These data processing steps showed that the single metastatic cancer patient's dataset, despite adaptive treatment decisions, could be rearranged to identify sufficient periods for treatment contrasts.

This study demonstrated the importance of clearly defining an estimand, or the parameter of interest,[32] to identify an interpretable causal effect in a single cancer patient's dataset. The steps of identifying an estimand, stating the causal structure using a DAG, and making explicit assumptions are commonly recommended in causal analysis to determine a treatment effect.[32,34-36] In this dataset, the estimand was defined as the average treatment effect conditional on the individual's baseline characteristics, where pre- and post-fulvestrant periods were compared under the FAS. The process of outlining the structure of the DAG showed that causal relationships between biomarkers, lesion size, and treatment status could be graphically represented, and an appropriate analytical method could be identified to address underlying assumptions such as time trends, carryover effects, and outcome-outcome effects. Together, these processes underscored a need for a more systematic approach to estimate causal treatment effects in personalized oncology.

Under the causal structure, a treatment effect was calculated as -38.1 (95% CI: -57.9, -18.4). This estimate suggested fulvestrant to be beneficial for this patient, as biomarker levels under the "always treated" scenario were lower than the "never treated" scenario. Although the result was statistically significant, the interval was relatively wide. The data-generating process may not have been fully accurate due to the model misspecification, or the number of biomarker and lesion measurements may have been limited. The calculated difference was found to be about 20% smaller than the observed difference, implying that the true effect of fulvestrant may not have been as large as anticipated. The difference in these results reveals that relying on simple pre/post mean differences may not be a sufficient method to estimate valid treatment effects, and a formal causal framework is needed to address the intricacies of treatment dynamics in oncology.

In this work, treatment exposure was described as a binary indicator at each time point to show whether a treatment had been administered at least once since the previous measurement. However, the exposure status did not include the exact dosage levels or drug decay dynamics over time. In addition, the original publication from which this dataset was derived reported multiple molecular profiling data of tumor biopsies at four distinct time points across the study course. Integrating molecular information that may underlie treatment resistance and an exposure status more clearly defined through dosage or decay information into causal frameworks is recommended to provide greater insights into underlying treatment response. Overall, a more comprehensive understanding of causal estimates in individual cancer settings may be made possible through combining clinical, molecular, and pharmacological data.

The development of biomarker tools for monitoring cancer treatment response has advanced substantially, allowing for longitudinal data collection that is less invasive and more timely.[37] For example, circulating tumor DNA can be used to monitor treatment response and reveal several treatment resistance mechanisms through liquid biopsies,[38,39] offering a less invasive alternative to tissue biopsies.[40] These advances will increase the sensitivity and specificity of cancer treatment responses,[41] although biomarkers alone will not be able to solve the challenge of treatment heterogeneity in cancer patients. More advanced tools that will allow adaptive data collection to be conducted more flexibly in individualized settings will be critical to increase the use of N-of-1 approaches in oncology. By combining these approaches with causal methods, N-of-1 designs have the potential to provide data to assist real-world treatment decisions, rather than merely summarizing results.

## 5.1 Recommendations for Causal Inference in Personalized Oncology

This study illustrated the possibility of applying causal inference methods in a single-cancer patient dataset, provided that data are collected at sufficient frequency, treatment periods are well defined, and a causal framework is used. Although this study focuses on a specific treatment period, the process of mapping treatments and outcomes under explicit causal assumptions provides a framework applicable in other settings with similar data. Treatment periods without intermediate changes, minimal gaps, and appropriate longitudinal follow-up were some key considerations for a meaningful comparison of treatment trajectories. Additionally, the results of data processing showed that treatment periods typically involved drug combinations. The use of combination therapies is often the norm in metastatic cancer[42] and indicates that future applications of causal inference in personalized oncology may need to shift from evaluating single agents to considering the joint effects of treatment combinations. After constructing a DAG to identify causal relationships and a target estimand, the need for a time-varying g-formula became clear, due to assumptions such as outcome–outcome effects, time trends, and carryover effects. This indicated that simply comparing different periods was insufficient to state as the causal effect. To ensure that treatment effect estimates are clinically meaningful, collecting well-defined data and applying a formal causal framework are recommended considerations for future applications of causal inference in personalized oncology. When feasible, the steps outlined in this study should be incorporated to assess causal treatment effects in single-patient settings.